\documentclass[twoside,a4paper,11pt]{sea9}
\usepackage{graphicx}
\usepackage{hyperref}
\usepackage{movie15}
\pdfoutput=1
\begin{document}
\pagenumbering{arabic}
\pagestyle{myheadings}
\thispagestyle{empty}
{\flushleft\includegraphics[width=\textwidth,bb=58 650 590 680]{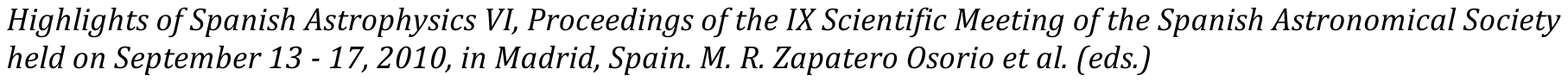}}
\vspace*{-2cm}
\begin{flushleft}
{\bf {\LARGE
%
The GaLAxy Cluster Evolution Survey (GLACE): introduction and first results
%
}\\
\vspace*{1cm}
%
M. S\'anchez-Portal$^{1,2}$,
J. Cepa$^{3,4}$, 
I. Pintos-Castro$^{4,3}$
R. P\'erez-Mart\'{\i}nez$^{1,2}$
I. Smail$^{5}$,
E. Alfaro$^{6}$, 
B. Altieri$^{1}$,
A. Arag\'on-Salamanca$^{7}$, 
C. Balkowski$^{8}$, 
M. Balogh$^{9}$,
A. Biviano$^{10}$, 
A. Bongiovanni$^{4,3}$, 
M. Bremer$^{11}$,
F. Castander$^{12}$, 
H. Casta\~neda$^{13}$,
N. Castro-Rodr\'{\i}guez$^{4,3}$,
D. Coia$^{1}$,
P.A. Duc$^{14}$, 
J. Geach$^{15}$,
I. Gonz\'alez-Serrano$^{16}$, 
C. Haines$^{17}$,
B. McBreen$^{18}$,
L. Metcalfe$^{1}$, 
I. P\'erez-Fourn\'on$^{4,3}$, 
A. M. P\'erez Garc\'{\i}a$^{4,3}$, 
B. Poggianti$^{19}$, 
J. M. Rodr\'{\i}guez-Espinosa$^{4,3}$, 
G. P. Smith$^{17}$,
S. Temporin$^{20}$
and 
I. Valtchanov$^{1}$
%
}\\
\vspace*{0.5cm}
%
$^{1}$
European Space Astronomy Centre (ESAC)/ESA, Madrid, Spain\\
$^{2}$
Ingenier\'{\i}a y Servicios Aeroespaciales, Madrid, Spain\\
$^{3}$
Universidad de La Laguna, Tenerife, Spain\\
$^{4}$
Instituto de Astrof\'{\i}sica de Canarias, La Laguna, Tenerife, Spain\\
$^{5}$
Institute for Computational Cosmology, Durham University, U.K.\\
$^{6}$
Instituto de Astrof\'{\i}sica de Andaluc\'{\i}a, CSIC, Granada, Spain\\
$^{7}$
School of Physics and Astronomy, University of Nottingham, U.K.\\
$^{8}$
GEPI, Observatoire de Paris \& CNRS,  Meudon, France\\
$^{9}$
Department of Physics and Astronomy, University of Waterloo, Canada\\
$^{10}$
INAF, Osservatorio Astronomico di Trieste, Italy\\
$^{11}$
Department of Physics, University of Bristol, U.K.\\
$^{12}$
Institut de Ci\`encies de l'Espai (CSIC), Barcelona, Spain\\
$^{13}$
Instituto Polict\'ecnico Nacional, M\'exico D.F., M\'exico\\
$^{14}$
Laboratoire AIM Saclay, CEA/IRFU, CNRS/INSU, Universit\'e Paris Diderot, France\\
$^{15}$
Department of Physics, McGill University, Montreal, Quebec, Canada\\
$^{16}$
Instituto de F\'{\i}sica de Cantabria, CSIC-Univ. de Cantabria, Santander, Spain\\
$^{17}$
School of Physics and Astronomy, University of Birmingham, U.K.\\
$^{18}$
University College, Belfield, Dublin 4, Ireland\\
$^{19}$
INAF, Osservatorio Astronomico di Padova, Italy\\
$^{20}$
Institute of Astro- and Particle Physics, University of Innsbruck, Austria
%
\end{flushleft}
%
\markboth{
The GLACE Survey: introduction and first results
}{ 
%
M. S\'anchez-Portal et al. 
%
}
\thispagestyle{empty}
\vspace*{0.4cm}
\begin{minipage}[l]{0.09\textwidth}
\ 
\end{minipage}
\begin{minipage}[r]{0.9\textwidth}
\section*{Abstract}{\small
\sf
%
Aimed at understanding the evolution of galaxies in clusters,  the GLACE survey is mapping a set of  
optical lines  ([O{\sc ii}]3727, [O{\sc iii}]5007, H$\beta$ and H$\alpha$/[N{\sc ii}] when possible) in several
galaxy clusters at z\,$\sim$\,0.40, 0.63 and 0.86, using the Tuneable Filters (TF) of the OSIRIS instrument \cite{Cepa05} 
at the 10.4m GTC telescope.
This study will address key
questions about the physical processes acting upon the infalling
galaxies during the course of hierarchical growth of clusters. 
GLACE is already ongoing: we present some preliminary results  
on our observations of the galaxy cluster Cl0024+1654 at z\,=\,0.395; GLACE@0.86 has been approved as ESO/GTC large project to be started in 2011.

%
\normalsize}
\end{minipage}
%
%
%
\section{Introduction \label{intro}}
The cores of nearby clusters are dominated by red
early-type galaxies, suggesting a long period of passive evolution
\cite{Bower98}. At higher redshifts, however, significant
evolution in galaxy properties is well-known through the Butcher-Oemler
(BO) effect \cite{BO84}: the increase in the fraction of blue galaxies in clusters at z\,$>$\,0.2.  
An equivalent increase in obscured star formation (SF) activity has also been seen in mid-IR surveys of
distant clusters (\cite{Coia05},  \cite{Geach06}) and equally an increasing population of AGN,
have been found in more distant clusters.  Even focusing on a single epoch, aspects of this same
evolutionary trend have been discovered in the outer parts of clusters
where significant
changes in galaxy properties can be clearly identified such as
gradients in typical colour or spectral properties with clustercentric
distance (\cite{Balogh99}, \cite{Pimbblet01}) and in the
morphology-density relation \cite{Dressler97}.
In a hierarchical model of structure formation, galaxies merge into
larger and larger systems as time progresses. 
It is quite likely that this
accretion process is responsible for a transformation of the properties
of cluster galaxies both as a function of redshift and as a function of
environment (\cite{Balogh00},  \cite{Kodama01}).
The central physical processes causing these evolutionary and
environmental changes remains elusive \cite{Treu03}. 
Possible
processes that have been proposed include: {\it (i)} Galaxy-ICM interactions: ram-pressure stripping, thermal evaporation of the ISM, 
turbulent and viscous stripping, pressure-triggered SF. When a slow decrease of the SF is produced, these mechanisms are collectively labelled as starvation. {\it (ii)} Galaxy-cluster gravitational potential interactions:  tidal compression, tidal truncation. {\it (iii)} Galaxy-galaxy interactions: mergers (low-speed interactions), harassment (high-speed interactions).  Nevertheless,
we do not yet understand whether the correlations of star-formation histories 
and large-scale structure are due to the advanced evolution 
in overdense regions, or to a direct physical effect on the star formation
capability of galaxies in dense environments \cite{Popesso07}.
This distinction can be made reliably if one has an accurate
measurement of star formation rate (or history), for galaxies spanning
a range of stellar mass and redshift, in different environments. 
The physical processes proposed above act on the star-forming (SF) and AGN population
within the clusters.  Although useful, optical
broad-band photometric surveys are too crude to reliably assign membership
or dynamics to individual galaxies, or to conclusively show that they were
SF.  Narrow-band imaging surveys are much more productive than spectroscopic ones
as a method to identify {\it all} the emission line galaxies (ELG) in a cluster.
Nevertheless, ``classical'' narrow-band imaging suffers
from ambiguity about the true fluxes of detected sources and does not provide dynamical
information about the population. 

\section{The GLACE project: objectives and description \label{objectives}}
The GaLAxy Cluster Evolution Survey (GLACE) is an innovative
survey of ELG and AGN in a well-studied and
well-defined sample of clusters in three narrow redshift windows, chosen to be 
relatively free of strong sky emission lines. This programme is undertaking a
panoramic census of the star formation and AGN activity within nine
clusters at z\,$\sim$0.40, 0.63 and 0.86, mapping a set of important lines:
H$\alpha$ (only at z\,$\sim$\,0.4), H$\beta$, [O{\sc ii}]3727 and [O{\sc iii}]5007; these maps of ELG are compared  with the
structures of these systems (as traced
by galaxies, gas and dark matter) to address 
several crucial issues: 

(1) Star formation in clusters: We will determine how the star formation properties of galaxies relate to
their position in the large scale structure. This will provide a key
diagnostic to test between different models for the environmental
influence on galaxy evolution. Each mechanism 
is most effective in a different environment (generally depending on the cluster-centric distance, e.g. \cite{Treu03}), leaving a footprint in
the data.  We are mapping the extinction-corrected star formation through H$\alpha$ and [O{\sc ii}], over a
large and representative region in a statistically useful sample of
clusters. The survey has been designed to reach SFR $\sim$2 M$_\odot$/yr (i.e. below
that of the Milky Way) with 1 magnitude of extinction at H$\alpha$\footnote{f$_{H\alpha}$\,=\,1.89\,$\times$\,10$^{-16}$\,erg\,s$^{-1}$ at z\,=\,0.4 using standard 
SFR$-$luminosity conversion factors \cite{Kennicut98}}.  Our first results
(see Section \ref{results}) indicate a good agreement with the requirement.  Another important question that can be addressed is related with the kind of
galaxies (type/mass) forming stars within clusters.  
The study of the morphology of confirmed ELGs allows investigating
 the connection between the truncation
of star formation and the morphological transformation of infalling
galaxies (e.g.\cite{Poggianti99}). In addition, by determining the total
integrated star formation rates, we can construct a
star-formation history for galaxies in clusters, as has
been done in the field (e.g. \cite{Madau98}).

(2) The role of AGNs:  
Whether the fraction of AGN is
environment-dependent or not is a matter of debate: while some results \cite{Miller03} point towards 
a lack of dependency of the AGN fraction with the local galaxy density, 
other authors  \cite{Kauffman04} conclude that high-luminosity AGN do avoid high-density
regions. Such a lack of AGN in clusters, as compared to the field, may
be related to the evolution of galaxies as they enter the cluster
environment.  On the other hand, claims for an enhanced
fraction of AGN in groups (eg \cite{Popesso06}) point at AGN-stimulating
processes such as galaxy--galaxy interactions or mergers that are
particularly effective in the low-velocity dispersion and (relatively)
high-density environments typical of groups (and filaments). A
complete census of the AGN population in clusters at different
redshifts, and in cluster regions characterized by different mean
densities and velocity dispersions, can help us to constrain the physics
behind the onset of the AGN activity in galaxies. Our survey is 
sensitive to the AGN population within clusters, discriminated
from pure star formation by means of standard diagnostics (e.g. BPT diagrams).

(3)  The study of the distribution of galaxy metallicities 
with cluster radii is another potentially powerful mean to investigate 
evolution within clusters. As galaxies fall into clusters and travel toward
the cluster center along their radial orbits, they interact with 
the intra-cluster medium (ICM) and other cluster galaxies, thus getting progressively 
stripped of their gas reservoir. This process is expected to influence their
metal abundances and possibly generate a metallicity gradient across a cluster.
Not much is yet known on the metallicities of emission-line galaxies in
clusters and how it varies with cluster-centric distances. GLACE will
allow to derive extinction-corrected metallicities using N2 \cite{Denicolo02}, R23 
\cite{Pagel79}, and O3N2 \cite{Alloin79}, where N2 and O3N2
will be used to break the R23 degeneracy and to assess possible 
differences between N and O abundances vs metallicity. 

The survey can address many other interesting topics: for instance the cluster accretion history can be traced
by studying the census of ELGs at  different cluster-centric distances.  
In addition,  the survey will provide an accurate assessment of cluster membership, without 
the need of a spectroscopic follow-up. 

Regarding the technical implementation, the GLACE survey applies the technique of TF tomography:
for each line, a set of images are taken 
through the OSIRIS TF, each image tuned at a different wavelength (equally spaced), 
so that a rest frame velocity range of several thousands km/s (6500 km/s for our first target) centred at the mean 
cluster redshift is scanned for the full TF field of view of 8\,arcmin in diameter. 
Additional images are taken to compensate for the blueshift of the wavelength from centre 
to the edge of the field of view.
Finally, for each pointing and wavelength tuned, 
three dithered exposures allow correcting for etalon diametric 
ghosts, using combining sigma clipping algorithms. 

The TF FWHM and sampling (i.e.: the wavelength interval between 
consecutive exposures)
at H$\alpha$ are of 1.2 and 0.6 nm, 
respectively, to allow deblending H$\alpha$ from [N{\sc ii}]$\lambda$6584.
with an accuracy better than 3$\%$. 
For the rest of the lines, the largest available TF FWHM, 2.0\,nm is applied, 
with sampling interval of 1.0\,nm. These 
parameters allow a continuum subtraction accuracy better than 2$\%$ 
and a photometric accuracy better than 6$\%$.
The same pointing
positions are observed at every emission line. We have required to cover $\simeq$2 Virial radii (some 4 Mpc) within the targeted clusters. This determines the
number of OSIRIS pointings (2 at 0.4 and 0.63, 1 at 0.85).


%


\section{Initial results \label{results}}

The first target chosen for the GLACE survey is the rich cluster Cl0024+1654 at z\,=\,0.395; this object has been comprehensively studied; public catalogues\footnote{\texttt{http://www.astro.caltech.edu/$\sim$smm/clusters/}} include photometric data for 73318 objects detected and extracted in the {\it HST} WFPC2 sparse mosaic covering $0.5\times0.5$\,degrees \cite{Treu03}  and in the ground-based CFHT CFH12k $BVRI$ and Palomar WIRC $JK_s$ imaging.  Visually determined morphological types are given for all objects brighter than I=22.5.  In addition, thousands of photometric and spectroscopic redshift estimates are available. The catalogue of spectroscopically confirmed objects within the field (including foreground, cluster and background sources) comprises 1632 sources (see \cite{Moran07} and references therein).
A narrow-band Subaru H$\alpha$ (plus $BVRI$ broadband) survey has been carried out \cite{Kodama04}, although this has an unknown selection function due to the narrow-band filter transmission, as well as lacking the velocity resolution and ability to deconvolve [N{\sc ii}]  contributions of our OSIRIS observations. 
The cluster field also has deep {\it Spitzer}/MIPS observations at 24$\mu$m \cite{Geach06}.

So far, two observing campaigns (GTC semesters 09B and 10A), allocating a total of 18 hours have been completed. The red TF has been used to observe the [O{\sc iii}]5007, H$\beta$ and H$\alpha$/[N{\sc ii}]  lines\footnote{The [O{\sc ii}]3727 line will be observed when the blue TF and corresponding order sorting filters become available}. Here we present the first results on the processing of the H$\alpha$/[N{\sc ii}]  complex (5.15 hours of on-source integration time, covering the 9047--9341\,\AA\  range in 50 scan steps).  Data reduction was performed using a version of the TFRED package \cite{Jones02} modified for OSIRIS by A. Bongiovanni and private IDL scripts. Details of the processing are given in \cite{Pintos10}.

\begin{figure}
\center
\includegraphics[scale=0.37]{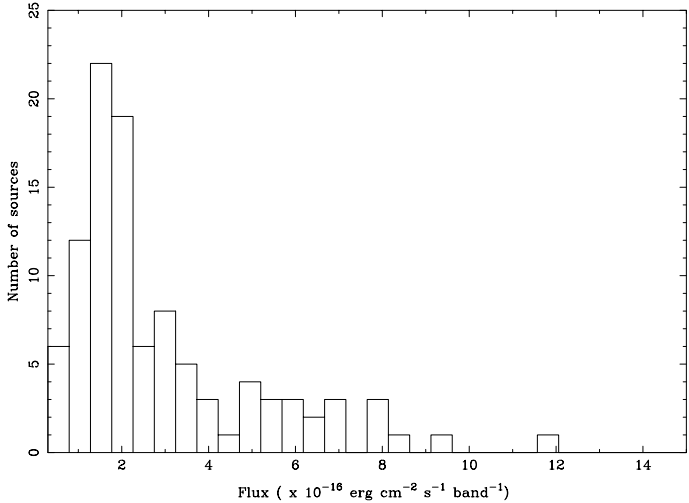}
\vspace{-0.75cm}
\caption{\label{flux_histo} {\small Flux histogram of the sub-sample of ELGs (103 sources). The fluxes represented here correspond to the maxima of the pseudo-spectra (i.e. line + continuum fluxes integrated within the largest flux single filter passband of 12\AA\  FWHM).}
}
\end{figure}

From our H$\alpha$/[N{\sc ii}] maps we have obtained a raw catalog of 1076 sources. In a first analysis, we have extracted 103 very robust (i.e. high S/N) emission line galaxies (ELG; comprise both star-forming galaxies and AGN). The completeness limit of the ELG sample (Fig.~\ref{flux_histo}) is $\sim$\,1.5\,$\times$\,10$^{-16}$\,erg\,s$^{-1}$cm$^{-2}$band$^{-1}$ (line + continuum) , i.e. very likely well within the GLACE requirements.

\begin{figure}
\center
\includegraphics[scale=0.7]{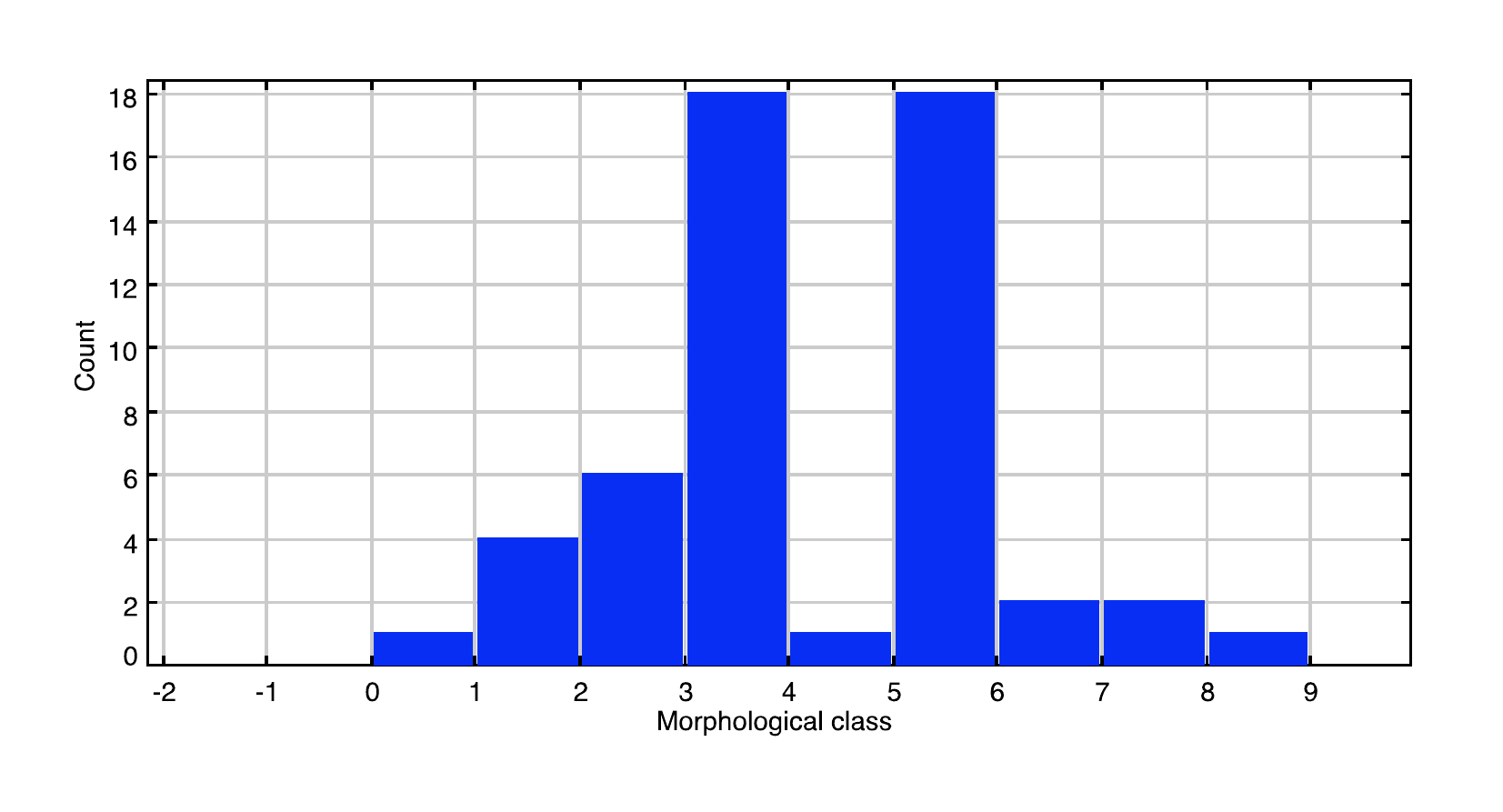}
\vspace{-0.75cm}
\caption{\label{morpho} {\small Morphological classification of the robust ELG detections, according to \cite{Abraham06}: \mbox{-2\,=\,star}, -1\,=\,compact, 0\,=\,E, 1\,=\,E/S0, 2\,=\,S0, 3\,=\,Sab, 4\,=\,S, 5\,=\,Scdm, 6\,=\,Irr, 7\,=\,peculiar, 8\,=\,merger, 9\,=\,defect
}}
\end{figure}

We have cross-matched our ELG catalogue with the public photometric and spectroscopic catalogues described above (see also \cite{Moran07}) by means of the Starlink TOPCAT tool using a search radius of 1.5\,arcsec, finding 96 matches (93\% of the sample). From these,  a morphological type was available for 53 objects (Fig.~\ref{morpho}). As expected, most of the ELGs are spiral galaxies (37), with a minor fraction of irregular, peculiar or mergers (5 objects). However, a significant fraction of the ELGs (10 sources) are early-type galaxies. 

\begin{figure}
\center
\includegraphics[scale=0.7]{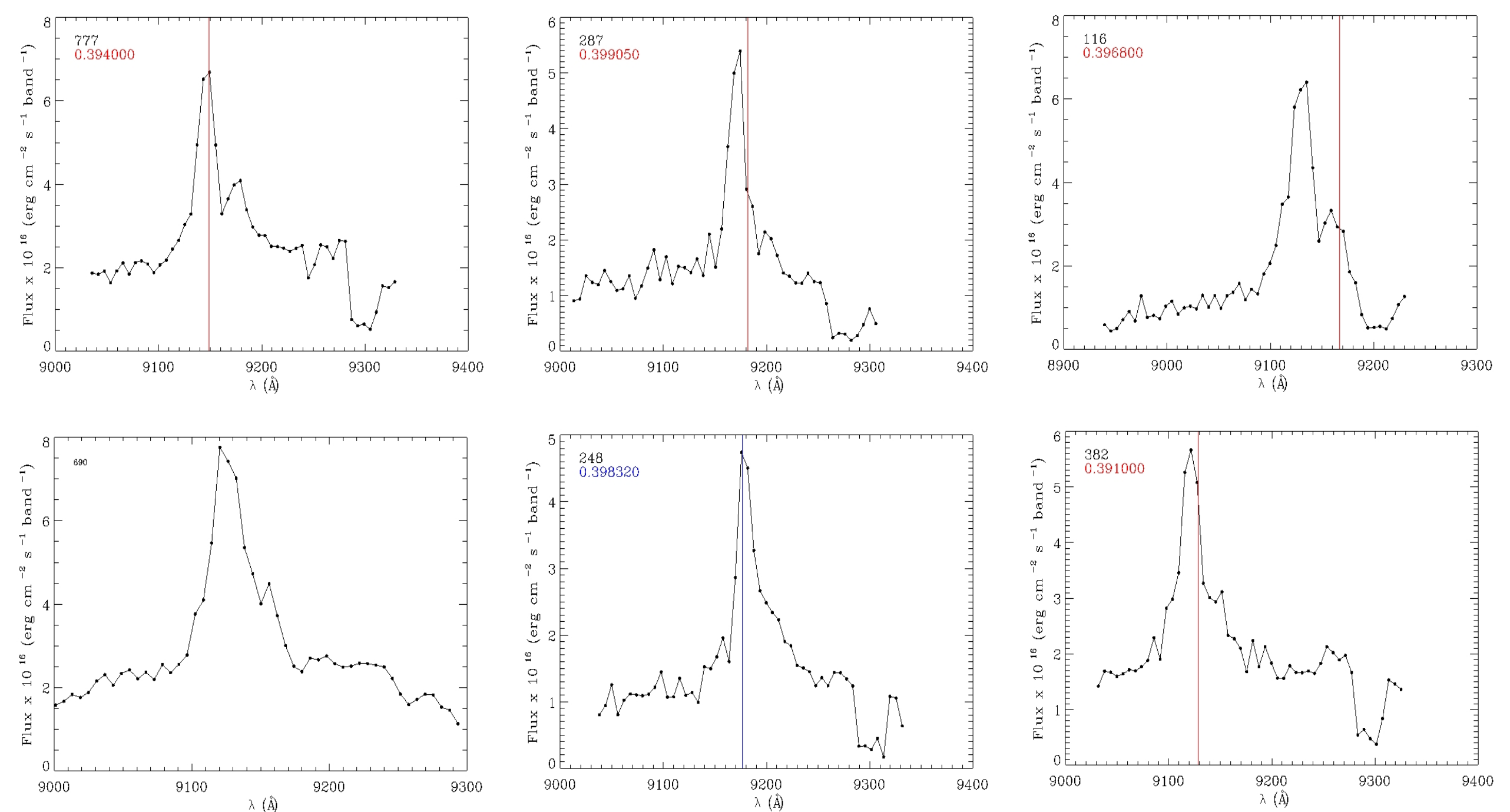}
\vspace{-0.75cm}
\caption{\label{spectra} {\small Examples of pseudo-spectra built from the H$\alpha$ scan (50 steps). Generally one of the [N{\sc ii}] doublet components is clearly resolved. When available, redshifts from the public catalogue have been included (blue--spectroscopic redshift from Keck/DEIMOS,  red--spectroscopic redshift from other sources). 
}}
\end{figure}

Redshift estimates were available for 90 of our ELGs. From these, a vast majority (83) have a redshift close to that of the cluster. There was a remarkably good agreement between the redshift estimates derived from the position of the H$\alpha$ line within our pseudo-spectra (see some examples in Fig.~\ref{spectra}) and that derived from spectroscopic measurements. Work in progress includes a precise deblending of the H$\alpha$ and [N{\sc ii}] lines and an accurate determination of the line wavelengths (the maxima in the pseudo-spectra plotted correspond to the central wavelength of the scan including the actual line maximum. This has therefore an uncertainty of $\pm$\,3\AA). This information will in turn allow us to establish cluster membership, galaxy dynamics, star formation rate and/or AGN nature of the emission.

\begin{figure}
\center
\includegraphics[scale=0.5]{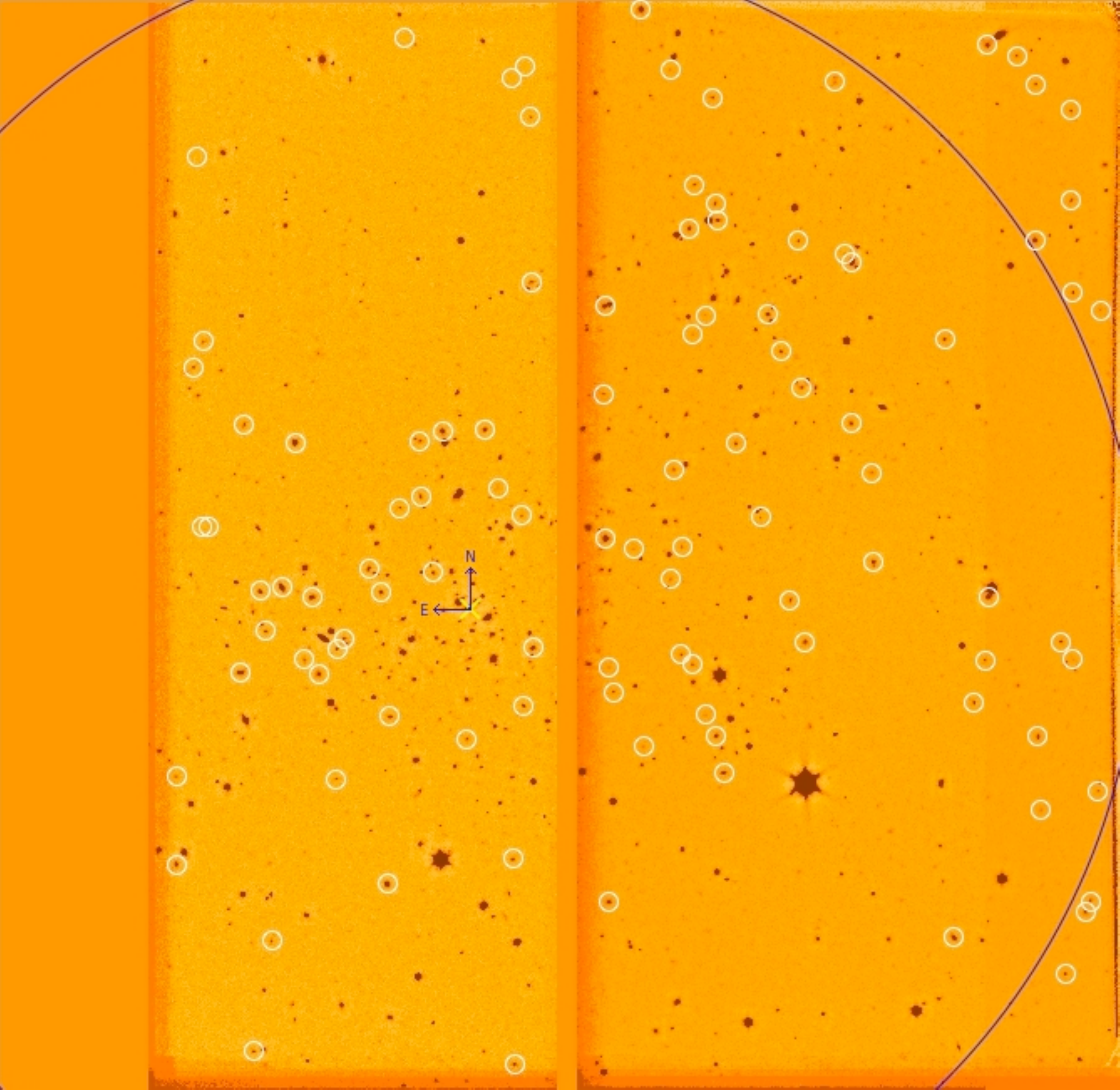}
\vspace{-0.25cm}
\caption{\label{deep_cat} {\small Deep cluster image obtained by adding up all the scan steps.  The light circles enclose  robust ELG detections.  The cross symbox ($\times$) marks the cluster centre. The large blue circle corresponds to r$_{vir}$.  }
}
\end{figure}

The spatial distribution of the ELG sample is depicted in Fig.~\ref{deep_cat} and maps the presence of two components: {\it (i)} a structure assembling onto the cluster core from the NW with an orientation almost in the plane of the sky. This structure has been already reported by other authors (\cite{Moran07}; \cite{Zhang05};  \cite{Kneib03}). 
 {\it (ii)} An infalling group at high velocity nearly along the line of sight to the cluster centre \cite{Moran07}.

%
%
\small  
%
\section*{Acknowledgments}   
%
This work was supported by the Spanish Plan Nacional de
Astronom\'\i a y Astrof\'\i sica under grant AYA2008--06311--C02--01.
We acknowledge support from the Faculty of the European Space Astronomy Centre (ESAC).
Based on observations made with the Gran Telescopio Canarias (GTC), installed in the Spanish Observatorio del Roque de los
Muchachos of the Instituto de Astrof\'\i sica de Canarias, in the island of La Palma. 
%

%

\begin{thebibliography}{}
\small
%
\bibitem{Alloin79}{Alloin, D. et al. 1979, A\&A, 78, 200}
\bibitem{Abraham06}{Abraham, R., et al., 1996, ApJS, 107, 1}
\bibitem{Balogh99}{Balogh, M., et al., 1999, ApJ, 527, 54}
\bibitem{Balogh00}{Balogh M., et al., 2000, ApJ, 540, 113}
\bibitem{Bower98}{Bower, R.G., et al., 1998, MNRAS, 299, 1193}
\bibitem{BO84}{Butcher, H. \& Oemler, A., 1984, ApJ, 285, 426}
\bibitem{Cepa05}{Cepa, J., et al., 2005, RevMexAA, 24, 1}
\bibitem{Coia05}{Coia, D. et al., 2005, A\&A, 430, 59}
\bibitem{Denicolo02}{Denicol\'o, G. et al. 2002, MNRAS, 330, 69}
\bibitem{Dressler97}{Dressler, A., et al., 1997, ApJ, 490, 577}
\bibitem{Geach06}{Geach, J., et al., 2006, ApJ, 649, 661}
\bibitem{Jones02}{Jones, D. H.  et al. 2002, MNRAS 329, 759}
\bibitem{Kauffman04}{Kauffman, G., et al., 2004, MNRAS, 353, 713}
\bibitem{Kennicut98}{Kennicutt, R.C., 1998, ARA\&A, 36, 189}
\bibitem{Kneib03}{Kneib, J.-P., et al., 2003, ApJ, 598, 804}
\bibitem{Kodama01}{Kodama, T. \& Bower, R., 2001,  MNRAS, 321, 18}
\bibitem{Kodama04}{Kodama, T., et al., 2004, MNRAS, 354, 1103}
\bibitem{Madau98}{Madau, P., et al., 1998, ApJ, 498, 106}
\bibitem{Miller03}{Miller, C.J., et al., 2003, ApJ, 597, 142}
\bibitem{Moran07}{Moran, S. et al., 2007, ApJ, 671, 1503}
\bibitem{Pagel79}{Pagel B.E.J., et al. 1979, MNRAS, 189, 95}
\bibitem{Pimbblet01}{Pimbblet, K.A., et al., 2001, MNRAS, 327, 588}
\bibitem{Pintos10}{Pintos-Castro, I., et al., 2010, these proceedings}
\bibitem{Poggianti99}{Poggianti, B.,et al., 1999, ApJ, 518, 576}
\bibitem{Popesso06}{Popesso, P. \& Biviano, A., 2006, A\&A, 460, L23}
\bibitem{Popesso07}{Popesso, P., et al., 2007, A\&A, 461, 411}
\bibitem{Treu03}{Treu, T., et al., 2003, ApJ, 591, 53}
\bibitem{Zhang05}{Zhang, Y.-Y., et al., 2005, A\&A, 429, 8}
%
%
\end{thebibliography}
\end{document}